\documentclass[a4paper]{jpconf}

\usepackage{graphicx}
\usepackage{wasysym}

\newcommand{\cA}{{\cal A}}  \newcommand{\cB}{{\cal B}}

  \newcommand{\cP}{{\cal P}}

  \newcommand{\cV}{{\cal V}}

\newcommand{\dd}{{\rm d}}

\begin{document}

\begin{flushright}
IFT-UAM/CSIC-11-75\\
UAB-FT-698
\end{flushright}

\title{Gravitational Anomaly and Hydrodynamics}

\author{Karl Landsteiner$^{1}$, Eugenio Meg\'{\i}as$^{1,2}$, Luis Melgar$^{1}$, Francisco Pena-Benitez$^{1,3}$}

\address{$^{1}$Instituto de F\'{\i}sica Te\'orica UAM/CSIC, C/ Nicol\'as Cabrera 13-15, Universidad Aut\'onoma de Madrid, Cantoblanco E-28049 Madrid, Spain} 
\address{$^{2}$Grup de F\'{\i}sica Te\`orica and IFAE, Departament de F\'{\i}sica, Universitat Aut\`onoma de Barcelona, Bellaterra E-08193 Barcelona, Spain}
\address{$^{3}$Departamento de F\'{\i}sica Te\'orica,
  Universidad Aut\'onoma de Madrid, Cantoblanco E-28049 Madrid, Spain}

\ead{karl.landsteiner@csic.es, emegias@ifae.es, luis.melgar@estudiante.uam.es, fran.penna@uam.es}

\begin{abstract}
We study the anomalous induced current of a vortex in a relativistic
fluid via the chiral vortical effect, which is analogous to the
anomalous current induced by a magnetic field via the chiral magnetic
effect. We perform this analysis at weak and strong coupling. We
discuss inequivalent implementations to the chemical potential for an
anomalous symmetry. At strong coupling we use a holographic model with
a pure gauge and mixed gauge-gravitational Chern-Simons term in the
action. We discuss the holographic renormalization and show that
the Chern-Simons terms do not induce new divergences.  Strong and weak
coupling results agree precisely. We also point out that the
holographic calculation can be done without a singular gauge field
configuration on the horizon of the black hole.
\end{abstract}

\section{Introduction}

Anomalies are responsible for the breakdown of a classical symmetry
due to quantum effects. In vacuum the anomaly appears as the
non-conservation of a classically conserved current in a triangle
diagram with two additional currents. In four dimension two types of
anomalies can be distinguished according to whether only spin one
currents appear in the triangle \cite{Adler:1969gk,Bell:1969ts} or if
also the energy-momentum tensor participates
\cite{Delbourgo:1972xb,Eguchi:1976db}. We will call the first type of
anomalies simply chiral anomalies and the second type gravitational
anomalies. In four dimension we should actually talk of mixed
gauge-gravitational anomalies since triangle diagrams with only
energy-momentum insertions are perfectly conserved (see
e.g.~\cite{AlvarezGaume:1983ig}). In a basis of only left-handed
fermions transforming under a symmetry generated by $T_A$ the presence
of chiral anomalies is detected by the non-vanishing of $d_{ABC} =
\frac 1 2\mathrm{Tr}(T_A\{T_B,T_C\})$ whereas the presence of a
gravitational anomaly is detected by the non-vanishing of $b_A=
\mathrm{Tr}(T_A)$.

Some studies showed that at finite temperature and density, anomalies
give rise to new non-dissipative transport phenomena in the
hydrodynamics of charged relativistic fluids. In particular magnetic
fields in the fluid induce currents via the so-called chiral magnetic
effect~\cite{Fukushima:2008xe,Yee:2009vw,Rebhan:2009vc}. Later studies
showed that a vortex in a fluid induces also a current parallel to the
axial vorticity vector~\cite{Erdmenger:2008rm,Son:2009tf}. On the
basis of linear response theory, hydrodynamic transport coefficients
can be extracted from the long-wavelength and low-frequency limits of
some retarded Green functions. This leads to the so called Kubo
formulas. For the chiral magnetic effect the Kubo formula has been
derived in~\cite{Kharzeev:2009pj,Hou:2011ze}. In~\cite{Amado:2011zx}
it was shown that the chiral vortical conductivity for charge and
energy transport can be obtained respectively from the retarded Green
functions
\begin{eqnarray}
\label{eq:sigmaV1}\sigma^\cV  &=& \lim_{k_c\rightarrow 0} \frac{i}{2k_c} \sum_{a,b}\epsilon_{abc}
\langle J^a T^{0b} \rangle|_{\omega=0} \,, \qquad \sigma^{\epsilon,\cV}  = \lim_{k_c\rightarrow 0} \frac{i}{2k_c}\sum_{a,b}\epsilon_{abc}
\langle T^{0a} T^{0b} \rangle|_{\omega=0}  \,,
 \end{eqnarray}
where $J^i$ is the (anomalous) current and $T^{ij}$ is the energy-momentum tensor~(see also~\cite{Landsteiner:2011iq,Landsteiner:2011tg} for details). 

In this manuscript we try to understand the effects anomalies have on
the transport properties of relativistic fluids, both in the weak and
strong coupling regimes, with special emphasis on the gravitational
anomaly.  Anomalies are very robust features of quantum field theories
and do not depend on the details of the interactions.  Therefore a
non-interacting theory at weak coupling is sufficient for our purpose
even without specifying to which gauge theory it corresponds to. By
the same way a rather general model that implements the correct
anomaly structure in the gauge-gravity setup is enough.  Our approach
for the latest case will therefore be a ``bottom up'' approach in
which we simply add appropriate Chern-Simons terms that reproduce the
relevant anomalies to the Einstein-Maxwell theory in five dimensions
with negative cosmological constant.

\section{Weak Coupling}
\label{sec:weak_coupling}

In this section we compute the anomalous transport coefficients in a
theory of free right-handed fermions~$\Psi^f$ transforming under a
global symmetry group $G$ generated by matrices~$(T_A)^f\,_g$.  The
chemical potential for the fermion $\Psi^f$ is given by $\mu^f= \sum_A
q_A^f \mu_A$, where we write the Cartan generator $H_A =
q_A^f\delta^f\,_g$ in terms of its eigenvalues, the charges $q_A^f$.
We define the chemical potential through boundary conditions on the
fermion fields around the thermal circle \cite{Landsman:1986uw} 
\begin{equation}\label{eq:bcs}
\Psi^f(\tau) = - e^{\beta \mu^f} \Psi^f(\tau-\beta) \,,
\end{equation} 
with
$\beta=1/T$. The currents can be expressed in terms of Dirac fermions
as
\begin{eqnarray}
J^i_A &=& \sum_{f,g=1}^N T_A^g\,_f \bar\Psi_g \gamma^i \cP_+ \Psi^f \,,\qquad T^{0i} =  \frac i 2 \sum_{f=1}^N\bar\Psi_f  ( \gamma^0  \partial^i + \gamma^i
\partial^0  ) \cP_+\Psi^f\,, \label{eq:JE}
\end{eqnarray}
where we used the chiral projector $\cP_\pm = \frac 1 2 (1\pm\gamma_5)$.
The fermion propagator is
\begin{eqnarray}
S(q)^f\,_g &=&  \frac{\delta^f\,_g}{2} \sum_{t=\pm}
\Delta_t(i\tilde\omega^f,\vec{q}) \cP_+ \gamma_\mu \hat q^\mu_t \,,\qquad \Delta_t( i\tilde\omega^f, q) = \frac{1}{i\tilde\omega^f - t E_q}\,,
\end{eqnarray}
with  $i\tilde\omega^f = i\tilde\omega_n + \mu^f$, $\hat q_t^\mu = (1, t \hat
q)$, $\hat{q} = \frac{\vec{q}}{E_q}$, $E_q=|\vec q |$ and $\tilde\omega_n=\pi T(2n+1)$ are the fermionic Matsubara frequencies. We can easily include left-handed fermions as well.

\subsection{Vortical conductivity}
\label{subsec:vortical_conductivity}

\begin{figure}[htb]
\begin{center}
\includegraphics[width=14pc]{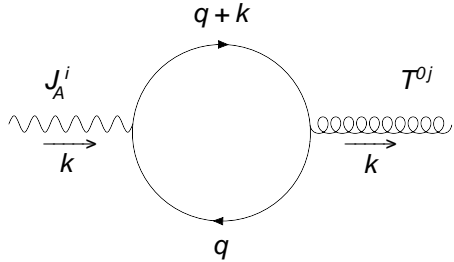}
\caption{1 loop diagram contributing to the vortical conductivity
Eq.~(\ref{eq:Ga}).}
\label{fig:1loop}
\end{center}
\end{figure}

 The vortical conductivity is defined from the retarded correlation
 function of the current $J^i_A(x)$ and the energy momentum tensor or
 energy current $T^{0j}(x^\prime)$, cf. Eq.~(\ref{eq:JE}), i.e.
\begin{equation}
G_A^\cV(x-x^\prime) = \frac{1}{2} \epsilon_{ijn}\,i \, \theta(t-t^\prime) 
\,\langle [J^i_{A}(x),T^{0j}(x^\prime)] \rangle \,. \label{eq:Ga}
\end{equation}
Going to Fourier space one gets the 1-loop contribution shown in Fig.~\ref{fig:1loop}.
The result for the zero frequency, zero momentum, vortical conductivity writes~\cite{Landsteiner:2011cp}
\begin{eqnarray}
(\sigma^\cV)_A &=& \frac{1}{8\pi^2} d_{ABC}\,\mu^B \mu^C  + \frac{T^2}{24} b_A  \,. \label{eq:sigmaV2}
\end{eqnarray}
The term involving the chemical potentials is induced by the chiral
anomaly. More interesting is the term $\sim T^2$ with a coefficient
that coincides with the gravitational anomaly
coefficient~\cite{AlvarezGaume:1983ig}. This means that a non-zero
value of this term have to be attributed to the presence of a
gravitational anomaly. Left handed fermions contribute in the same way
but with a relative minus sign.

\subsection{Magnetic conductivity}
\label{subsec:magnetic_conductivity}

The magnetic conductivity in the case of a vector and an axial $U(1)$ symmetry
was computed at weak coupling in~\cite{Kharzeev:2009pj}. The corresponding Kubo formula involves the two point function of the current. Following the same method, we get the result for a general symmetry group
\begin{equation}
(\sigma^\cB)_{AB} = \frac{1}{4\pi^2} d_{ABC}\, \mu^C \,.
\label{eq:sigmaB1}
\end{equation}
No contribution proportional to the gravitational anomaly coefficient is found in this case.

\section{Chemical potentials for anomalous symmetries}
\label{sec:chemical_potentials}

Before we go to how these results can be obtained at strong coupling
from a gauge-gravity duality we stop for a moment and reflect on the
formalism necessary to introduce the chemical potential.  Note that we
have been quite specific in how we introduced the chemical potentials
in the weak coupling calculation in (\ref{eq:bcs}). Furthermore this is
not the way chemical potentials are commonly discussed in
textbooks. Rather than demanding twisted boundary conditions on the
thermal circle it is far more common to consider a deformation of the
Hamiltonian
\begin{equation}\label{eq:defomredH}
 H \rightarrow H - \mu Q \,,
\end{equation}
where $Q$ is the charge in question. We can think of this as arising
from the coupling of a (fiducial) gauge field $A_\mu$ to the current
$j^\mu$ of the form $\int d^4x A_\mu j^\mu$ and giving a vacuum
expectation value to $A_0 = \mu$. With the fiducial gauge field we
have gauge invariance now and we can remove of course the $\mu Q$
coupling in the Hamiltonian by the gauge transformation $A_0
\rightarrow A_0 + \partial_0 \chi$ with $\chi = -\mu t$. Along the
imaginary time direction $t= -i \tau$ this introduces of course just
the twist in the boundary conditions on the fields in
(\ref{eq:bcs}). As long as we have honest non-anomalous symmetries
under consideration we have therefore two (gauge)-equivalent
formalisms of how to introduce the chemical potential summarized in table~\ref{tab:chem_pot} \cite{Evans:1995yz}.
\begin{table}
\caption{\label{tab:chem_pot}Two formalisms to chemical potential.}
\begin{center}
\begin{tabular}{lll}
\br
Formalism & Hamiltonian & Boundary conditions  \\ 
\mr
(A) & $H-\mu Q$ &  $\Psi(\tau) = -  \Psi(\tau-\beta)$\\ 
(B) & $H$ & $\Psi(\tau) = - e^{\beta \mu} \Psi(\tau-\beta)$ \\
\br
\end{tabular}
\end{center}
\end{table}
One convenient point of view on formalism (B) is the following. In a
real time Keldysh-Schwinger setup we demand some initial conditions at
initial (real) time $t=t_i$. These initial conditions are given by the
boundary conditions in (B). From then on we do the (real) time
development with the microscopic Hamiltonian $H$. This seems an
especially suited approach to situations where the charge in question
is not conserved by the real time dynamics. In the case of an
anomalous symmetry we can start at $t=t_i$ with a state of certain
charge but this charge does indeed decay over (real) time due to
non-perturbative processes (instantons) or at finite temperture due to
thermal sphaleron processes \cite{Moore:2010jd}. These processes are
however supressed at large $N$ and so can not be seen easily in the
gauge-gravity correspondence.  Taking this as excuse we simply ignore
them, but keep them in the back of our head as motivation for favoring
formalism (B) in the case of an anomalous symmetry, to which we come
right now.

Let us assume now that $Q$ is an anomalous charge and start with our
favoured formalism (B). We ask what happens if we do now the gauge
transformation that would bring us to formalism (A). Since the
symmetry is anomalous this means that the action transforms as
\begin{equation}
S[A+\partial\chi] = S[A] + \int d^4x\, \chi \epsilon^{\mu\nu\rho\lambda}\left(C_1 F_{\mu\nu}F_{\rho\lambda} + C_2 R^\alpha\,_{\beta\mu\nu} R^\beta\,_{\alpha\rho\lambda}\right) \,,
\end{equation}
with the anomaly coefficients $C_1$ and $C_2$ depending on the chiral
fermion content. It follows that formalisms (A) and (B) are physically
inequivalent now, because of the anomaly. However, we would like to
still come as close as possible to the formalism of (A) but in a form
that is physically equivalent to the formalism (B). To achieve this we
proceed by introducing a non-dynamical axion field $\Theta(x)$ and the
vertex
\begin{equation}
S_\Theta[A,\Theta] = \int d^4x \, \Theta\epsilon^{\mu\nu\rho\lambda}\left(C_1 F_{\mu\nu}F_{\rho\lambda} + C_2 R^\alpha\,_{\beta\mu\nu} R^\beta\,_{\alpha\rho\lambda}\right) \,. \label{eq:Saxion}
\end{equation}
If we demand now that the ``axion'' transforms as $\Theta \rightarrow \Theta - \chi$ under gauge transformations
we see that the action
\begin{equation}
S_{tot}[A,\Theta] = S[A] + S_\Theta[A,\Theta] \label{eq:Stot} 
\end{equation}
is gauge invariant. Note that this does not mean that the theory is
not anomalous now. We introduce it solely for the purpose to make
clear how the action has to be modified such that two field
configurations related by a gauge transformation are physically
equivalent. In other words $\Theta$ is a coupling and not a field. The
gauge field configuration that corresponds to formalism (B) is simply
$A_0=0$. A gauge transformation with $\chi=\mu t$ on the gauge
invariant action $S_{tot}$ makes clear that a physically equivalent
theory is obtained by chosing the field configuration $A_0=\mu$ and
the {\em coupling} $\Theta = -\mu t$. If we define the current through
the variation of the action with respect to the gauge field we get an
additional contribution from $S_\Theta$,
\begin{equation}
j_\Theta^\mu = 4 C_1 \epsilon^{\mu\nu\rho\lambda} \partial_\nu \Theta F_{\rho\lambda}\,, 
\end{equation}
and evaluating this for $\Theta= -\mu t$ we get the spatial current
\begin{equation}\label{eq:jTheta}
j_\Theta^m = 4 C_1 \mu B_m\,,
\end{equation}
(note that $\epsilon^{0ijk} = -\epsilon_{ijk}$ in Minkowski
space). This is {\em not} the chiral magnetic effect! This is only the
contribution to the current that comes from the new coupling that we
are forced to introduce by going to formalism (A) from (B) in a
(gauge)-equivalent way. The chiral magnetic and vortical effect are on
the contrary non-trivial results of dynamical one-loop calculations as
shown in the previous section. What is the Hamiltonian now based on
the modified formalism (A)? We have to take of course the new coupling
generated by the non-zero $\Theta$. The Hamiltonian now is
therefore\footnote{This form of the Hamiltonian (without the
  gravitational part) is the one advocated in~\cite{Rubakov:2010qi}.}
\begin{equation}
H - \mu \left( Q + 4 \int d^3x\, (C_1 \epsilon^{0ijk} A_i \partial_j A_k+  C_2 K^0 )\right)\,, 
\end{equation}
where $K^0$ is the zero component of the graviational Chern-Simons current 
$K^\mu = \epsilon^{\mu\nu\rho\lambda} \Gamma^\alpha_{\beta\nu}\left( \partial_\rho \Gamma^\beta_{\alpha\lambda} + \frac 2 3 \Gamma^\beta_{\rho\sigma} \Gamma^\sigma_{\alpha\lambda} \right)$, fulfilling
$\partial_\mu K^\mu = \frac 1 4 \epsilon^{\mu\nu\rho\lambda}R^\alpha\,_{\beta\mu\nu} R^\beta\,_{\alpha\rho\lambda}$.

\begin{figure}[htb]
\begin{center}
\includegraphics[scale=1]{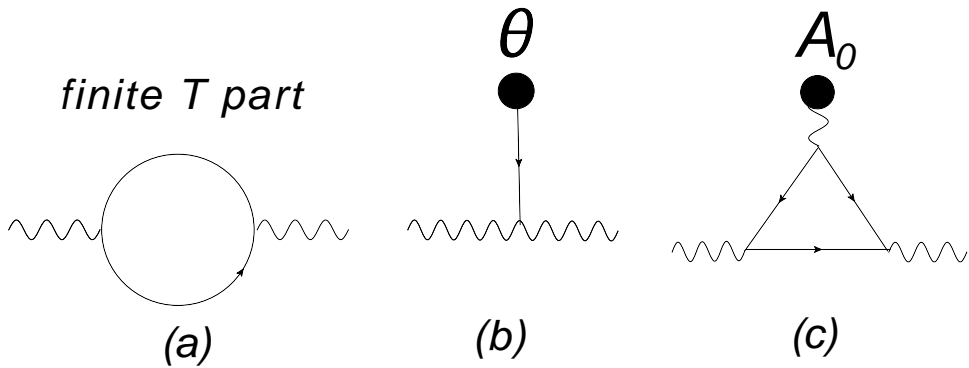}
\caption{The three different contributions to the current current correlator relevant for the
calculation of the chrial magnetic effect. (a) is the (UV-finite) finite
temperature part, (b) is the part stemming from the (non-dynamical) axion vertex and (c) stems
from the vacuum (T=0) triangle graph coupled to the background gauge field.}
\label{fig:Contribs}
\end{center}
\end{figure}

Notice that in Formalism (A) we really have three contributions now to
the current. One is formally a tree level contribution, namely the
current (\ref{eq:jTheta}), but there are also two different one-loop
contributions. The first one comes from the UV-finite part of a finite
temperature two point function.  This is the part that is typically
considered in the weak coupling Kubo formula calculations in
\cite{Kharzeev:2009pj, Landsteiner:2011cp}.  But in formalism (A) we
also have a (formally UV-divergent) contribution from the vacuum
triangle diagram that couples to the gauge-field background! This has
been first pointed out and emphasized
in~\cite{Gynther:2010ed}. Graphically in formalism (A) we have to sum
the graphs (a), (b) and (c) of figure~\ref{fig:Contribs}.  With
$\Theta = -\mu t$ and $A_0=\mu$ the contributions (b) and (c) cancel
each other.

We could also consider the contribution of the axion vertex to the
energy-momentum tensor in the case when a gravitational anomaly is
present. Since the Riemann tensor is however of second order in
derivatives it is clear that the correspondent contribution to the
energy current, i.e. the $T^{0i}$ components of the energy-momentum
tensor, will be of third order in derivatives\footnote{This has
  recently been made explicit in \cite{Kimura:2011ef}.}.  In contrast
the finite temperature one-loop graphs in Figs.~\ref{fig:1loop}
and~\ref{fig:Contribs} are first order in derivatives and contribute
therefore to ordinary first order hydrodynamics!

Our point of view is that the true anomalous transport effects are
captured by the finite temperature graphs (a). These are also the only
ones that contributes in formalism (B).  We will apply these
considerations now to the holographic calculation.

\section{Strong Coupling}
\label{sec:strong_coupling}

We present in this section the computation of the anomalous transport coefficients at strong coupling within a holographic model in five dimensions including terms which conveniently mimics the chiral and gauge-gravitational anomalies.

\subsection{Holographic Model}
\label{sec:holo_model}

 Given an outward pointing normal vector $n^A \propto
 g^{AB} \frac{\partial r}{\partial x^B}$ to the holographic boundary
 of an asymptotically AdS space with unit norm $n_A n^A =1$, the
 induced metric takes the form~$h_{AB} = g_{AB}- n_A n_B$. The action
 of our model is defined by
\begin{eqnarray}
S &=& \frac{1}{16\pi G} \int d^5x \sqrt{-g} \left[ R + 2 \Lambda -
  \frac 1 4 F_{MN} F^{MN} \right.\\ &&\left.+ \epsilon^{MNPQR} A_M
  \left( \frac\kappa 3 F_{NP} F_{QR} + \lambda R^A\,_{BNP} R^B\,_{AQR}
  \right) \right] + S_{GH} + S_{CSK} \,,\\ S_{GH} &=& \frac{1}{8\pi G}
\int_\partial d^4x \sqrt{-h} \, K \,,\\ S_{CSK} &=& - \frac{1}{2\pi G}
\int_\partial d^4x \sqrt{-h} \, \lambda n_M \epsilon^{MNPQR} A_N
K_{PL} D_Q K_R^L \,, 
\end{eqnarray}
where $S_{GH}$ is the usual Gibbons-Hawking boundary term and
$D_A=h_A^B\nabla_B$ is the covariant derivative on the four
dimensional boundary. The second boundary term $S_{CSK}$ is needed if
we want the model to reproduce the gravitational anomaly at general
hypersurface.  The most important fact about this action is the
presence of two Chern-Simons terms. The one proportional to the parameter~$\kappa$ is a pure gauge field CS term and the one proportional to~$\lambda$ a mixed gauge-gravitational CS term. Both of them are diffeomorphism invariant, and they do depend however explicitly on the gauge connection $A_M$. The bulk
equations of motion are
\begin{eqnarray}\label{eqgrav}
 G_{MN} - \Lambda g_{MN} &=& \frac 1 2 F_{ML} F_N\,^L - \frac 1 8 F^2 g_{MN} + 2 \lambda \epsilon_{LPQR(M} \nabla_B\left( F^{PL} R^B\,_{N)}\,^{QR} \right) \,, \label{eq:Gbulk}\\\label{eqgauge}
\nabla_NF^{NM} &=& - \epsilon^{MNPQR} \left( \kappa F_{NP} F_{QR} + \lambda  R^A\,_{BNP} R^B\,_{AQR}\right) \,,  \label{eq:Abulk}
\end{eqnarray}
and they are gauge and diffeomorphism covariant. In~\ref{sec:holo_renorm} we discuss the holographic renormalization of the model within the Hamiltonian approach. This leads to the following counterterm of the action~\cite{Landsteiner:2011iq}
\begin{eqnarray}
S_{ct} &=& - \frac{(d-1)}{8\pi G} \int_\partial d^4x \sqrt{-\gamma} \bigg[
1 + \frac{1}{(d-2)}P \nonumber \\
&&\qquad\qquad- \frac{1}{4(d-1)} \left( P^i_j P^j_i - P^2 -  \frac{1}{4} \hat{F}_{(0)}\,_{ij} \hat{F}_{(0)}\,^{ij} \right)\log e^{-2r} \bigg] \,,
\end{eqnarray}
where $P = \frac{\hat{R}}{2(d-1)}$ and $P^i_j = \frac{1}{(d-2)} \left[
  \hat{R}^i_j - P \delta^i_j \right]$. As a remarkable fact there is
no contribution in the counterterm coming from the gauge-gravitational
Chern-Simons term. This means that this term does not induce new
divergences, and so the renormalization is not modified by it.

\subsection{Evaluation of Transport Coefficients}
\label{sec:trans_coeff}

The AdS/CFT dictionary  tells us how to compute the retarded propagators \cite{Son:2002sd,Herzog:2002pc}. Since we are interested in the linear response limit, we split the metric and gauge field into a background part and a linear perturbation,
\begin{equation}
g_{MN} = g^{(0)}_{MN} + \epsilon \, h_{MN} \,, \qquad  A_{M} = A^{(0)}_{M} + \epsilon \, a_{M} \, .
\end{equation}
Inserting these fluctuations-background fields in the action and expanding up to second order in $\epsilon$ one can read the second order action which is needed to get the desired propagators~\cite{Kaminski:2009dh}.  If we construct a vector $\Phi^I$ with the components of $a_M$ and $h_{MN}$ and Fourier transforming it, then it is possible to write the complete second order action on-shell as a boundary term
\begin{equation}
\delta S^{(2)}_{ren}=\int \frac{\dd^d k}{(2\pi)^d} \lbrace \Phi^I_{-k} \cA_{IJ} \Phi '^J_k + \Phi^I_{-k}  \cB_{IJ} \Phi^J_k \rbrace\Big{|}_{r\to\infty}\,.  \label{eq:2ndor}
\end{equation}
From~(\ref{eq:2ndor}) it is possible to compute the holographic
response functions by applying the prescription of
\cite{Son:2002sd,Herzog:2002pc,Kaminski:2009dh,Amado:2009ts}.  For a
coupled system the holographic computation of the correlators consists
in finding a maximal set of linearly independent solutions that
satisfy infalling boundary conditions on the horizon and that source a
single operator at the AdS boundary. To do so we construct a matrix of
solutions $F^I\,_J (k,r)$ such that each of its columns corresponds to
one of the independent solutions and normalize it to the unit matrix
at the boundary. Finally using this decomposition one obtains the
matrix of retarded Green functions
\begin{equation} 
G_{IJ}(k)= -2 \lim_{r\to\infty} \left(\cA_{IM}
(F^M\,_J (k,r))' +\cB_{IJ}\right)\,. \label{eq:GR}    
\end{equation}

The system of equations (\ref{eqgrav})-(\ref{eqgauge}) admit the
following exact background AdS Reissner-Nordstr\"om black-brane
solution 
\begin{eqnarray} 
\dd s^2 &=& \frac{\bar r^2}{L^2}\left(-f(\bar r) \dd t^2
+\dd \vec{x}^2\right)+\frac{L^2}{\bar r^2 f(\bar r)} \dd \bar
r^2\,,\qquad f(\bar r) = 1-\frac{M L^2}{\bar r^4}+\frac{Q^2 L^2}{\bar r^6}\,,  \\
A^{(0)} &=& \phi(\bar r)\dd t = \left(\alpha-\frac{\mu \,\bar r_{{\rm H}}^2}{\bar r^2}\right)\dd t \,,  
\end{eqnarray}
where the horizon of the black hole is located at $\bar r=\bar r_{\rm
  H}$.  We would like to draw attention to the boundary value of the
gauge field $\alpha$. We will make two choices, the choice (A) where
$\alpha=\mu$ and the choice (B) where $\alpha=0$. Note that with choice
(A) the gauge field vanished at the horizon!  At this point the reader
will not be too surprised by our claim that the choice (A) corresponds
to the formalism (A) of the previous section and choice (B) to the
formalism (B). We know now however, that if we want (A) to be
physically equivalent to (B) we also need to take into account the
pure boundary action $S_\Theta[A,\Theta]$. In the holographic setup
this is a {\em pure} boundary term, $\Theta$ does not extend into the
bulk and is therefore not dynamical. If we do not include this
coupling it is known that the choices (A) and (B) give different
results. In fact it has caused quite some confusion that the choice
(A) without taking into account the $\Theta$-coupling gives a
vanishing result for the chiral magnetic effect as shown in~\cite{
  Rebhan:2009vc}. The alternative approach (B) causes however some
uneasyness because now the gauge field necessarily is singular at the
horizon (though not the field strength), but it gives the correct
result for the chiral magnetic effect with no need of including the
pure boundary action $S_\Theta[A,\Theta]$, as it was done
in~\cite{Gynther:2010ed}. The nice feature of the modified approach
(A) in holography is that we can now happily make the gauge field
vanish at the horizon and still obtain the correct result for the
anomalous conductivities on the boundary~\smiley

The parameters $M$, $Q$ and Hawking temperature of the RN black hole write
\begin{eqnarray} 
M=\frac{\bar r_{\rm
    H}^4}{L^2}+\frac{Q^2}{\bar r_{\rm H}^2}\quad,\quad Q=\frac{\mu\,
  \bar r_{\rm H}^2}{\sqrt{3}}\,, \qquad T=\frac{\bar r_{\rm H}^2}{4\pi\, L^2} f'(\bar r_{\rm
  H}) = \frac{ \left(2\, \bar r_{\rm H}^2\, M - 3\, Q^2 \right)}{2
  \pi \,\bar r_{\rm H}^5} \,.
\end{eqnarray} 
We just consider momentum fluctuations of the fields transverse to the
perturbations of the momentum, i.e. we focus on the shear sector. Then
one arrives at a system of four second order differential
equations. The relevant physical boundary conditions on fields are:
$h^\alpha_t(0)=\tilde H^\alpha$, $B_\alpha(0)=\tilde B_\alpha$; where
the `tilde' parameters are the sources of the boundary operators.  The
second condition compatible with the ingoing one at the horizon is
regularity for the gauge field and vanishing for the metric
fluctuation \cite{Amado:2011zx}.

After solving the system of differential equations perturbatively
(see~\cite{Landsteiner:2011iq} for details), one can go back to the
formula (\ref{eq:GR}) and compute the corresponding holographic Green
functions. Finally, by using the Kubo formulas~(\ref{eq:sigmaV1}) one
recovers the conductivities
\begin{eqnarray}
\label{eq:sigb}\sigma^\cB &=& -\frac{\sqrt{3}\,  Q \, \kappa}{2 \pi
  \,G\, \bar r_{\rm{H}}^2} =  \frac{ \mu}{4 \pi^2}\,,\\
\label{eq:sigv}\sigma^\cV &=&\sigma^{\epsilon,\cB}=-\frac{3\, Q^2\,  \kappa }{4 \pi\,G\, \bar  r_{\rm{H}}^4}-\frac{2\lambda \pi T^2}{G}= \frac{\mu^2}{8\pi^2} +\frac{T^2}{24} \,, \\
\label{eq:sigve}\sigma^{\epsilon,\cV} &=&-\frac{ \sqrt{3}\, Q^3\,
  \kappa}{2\pi\, G\, \bar r_{\rm{H}}^6} - \frac{4\pi \sqrt{3}   Q T^2
  \lambda}{G \,\bar r_{\rm H}^2} =\frac{\mu^3}{12 \pi^2}+\frac{\mu
  T^2}{12}\,,
\end{eqnarray}
where we have included for completeness the chiral magnetic
conductivity~$\sigma^\cB$ and the magnetic conductivity for energy
current~$\sigma^{\epsilon,\cB}$.

In a hydrodynamic framework we can summarize our findings in the constitutive relations
\begin{eqnarray}
 T^{\mu\nu}  &=& (\epsilon+P) u^\mu u^\nu + P \eta^{\mu\nu} + Q^\mu u^\nu + Q^\nu u^\mu \,,\\
 J^\mu &=& n u^\mu + N^\mu\,,
\end{eqnarray}
with the first order in derivative terms
\begin{eqnarray}
 N^\mu &=& \sigma^{\cal B} {\cal B}^\mu + \sigma^{\cal V} \Omega^\mu \,,\\
Q^\mu &=& \sigma^{\epsilon,\cal B} {\cal B}^\mu + \sigma^{\epsilon,\cal V} \Omega^\mu\,.
\end{eqnarray}
For simplicity of the expressions we have dropped here the usual dissipative terms related to shear and bulk viscosities or
electric conductivity.
The equilibrium quantities $\epsilon, P, n$ are energy density, pressure and charge density and we
defined the (covariant) magnetic field ${\cal B}^\mu = \epsilon^{\mu\nu\rho\lambda} u_\nu \partial_\rho A_{\lambda}$ and vorticity vector $\Omega^\mu = \epsilon^{\mu\nu\rho\lambda} u_\nu \partial_\rho u_\lambda$.

According to the discussion above,
these values for the coefficients can be obtained, either by using the
action of the holographic model and setting the deformation parameter
$\alpha$ to zero (choice B), or considering the total
action~$S_{tot}[A,\Theta]$, Eq.~(\ref{eq:Stot}), and setting
$\alpha=\mu$ and $\Theta= -\mu t$ (choice A). In the latest case the
axion term~(\ref{eq:Saxion}) induces a contribution in
matrices~$\cA_{IJ}$ and $\cB_{IJ}$, see Eq.~(\ref{eq:2ndor}). The
expression for $\sigma^\cB$ is in perfect agreement with the
literature and the one for $\sigma^\cV$ shows the extra $T^2$ term
already predicted in~\cite{Landsteiner:2011cp} and shown in
Section~\ref{sec:weak_coupling}. In fact the numerical coefficients
coincide precisely with the ones obtained in weak coupling when
specifying Eqs.~(\ref{eq:sigmaV2}) and (\ref{eq:sigmaB1}) to the
$U(1)_R$ group. This we take as a strong hint that the anomalous
conductivities are indeed completely determined by the anomalies and
are not renormalized beyond one loop.

\section{Discussion and conclusion}
\label{sec:discussion}

In the presence of external sources for the energy momentum tensor and
the currents, the anomaly is responsible for a non conservation law of
the latter. This is conveniently expressed
through~\cite{AlvarezGaume:1983ig}
\begin{equation}
 \nabla_\mu J_A^\mu= \epsilon^{\mu\nu\rho\lambda}\left( \frac{d_{ABC}}{32\pi^2} 
F^B_{\mu\nu} F^C_{\rho\lambda} + \frac{b_A}{768\pi^2} 
R^\alpha\,_{\beta\mu\nu}
R^\beta\,_{\alpha\rho\lambda}\right) \,.\label{eq:anomaly}
\end{equation}
The axial and mixed gauge-gravitational anomaly coefficients are defined respectively by
\begin{eqnarray}
d_{ABC} &=& \frac 1 2 [ \tr( T_A \{ T_B, T_C\} )_R -  \tr( T_A \{ T_B, T_C\} )_L
]\,, \label{eq:chiralcoeff}\\
b_A &=& \tr (T_A)_R - \tr (T_A)_L\label{eq:gravcoeff} \,,
\end{eqnarray}
where the subscripts $R,L$ stand for the contributions of right-handed and
left-handed fermions.

We have computed the magnetic and vortical conductivity of a
relativistic fluid at weak coupling and we find contributions that are
proportional to the anomaly coefficients (\ref{eq:chiralcoeff}) and
(\ref{eq:gravcoeff}). Non-zero values of these coefficients are a
necessary and sufficient condition for the presence of anomalies
\cite{AlvarezGaume:1983ig}.  Therefore the non-vanishing values of the
transport coefficients~(\ref{eq:sigmaV2}) and (\ref{eq:sigmaB1}) have
to be attributed to the presence of chiral and gravitational
anomalies.

In order to perform the analysis at strong coupling via AdS/CFT
methods, we have defined a holographic bottom up model that implements
both anomalies in four dimensional field theory via gauge and mixed
gauge-gravitational Chern-Simons terms. We have computed the anomalous
magnetic and vortical conductivities from a charged black hole
background and have found a non-vanishing vortical conductivity
proportional to $\sim T^2$. These terms are characteristic for the
contribution of the gravitational anomaly and they even appear in an
uncharged fluid. The~$T^2$ behavior had appeared already previously
in neutrino
physics~\cite{Vilenkin:1980ft,Vilenkin:1980zv}. In~\cite{Neiman:2010zi}
similar terms in the vortical conductivities have been argued for
without any relation to the gravitational anomaly. However so far the
effects of gravitational anomalies have not been taken into account in
the purely hydrodynamic treatments, and therefore the $T^2$ terms
appear simply as undetermined integration constants. The numerical
values of the anomalous conductivities computed at strong coupling are
in perfect agreement with weak coupling calculations, and this
suggests the existence of a non-renormalization theorem including the
contributions from the gravitational anomaly.

So far we have computed the transport coefficients, and in particular their gravitational anomaly contributions, via Kubo formulas. It would be interesting to calculate directly the constitutive relations of the hydrodynamics of anomalous currents via the fluid/gravity correspondence within the model used in the present paper~\cite{Erdmenger:2008rm,Bhattacharyya:2008jc,Banerjee:2008th}. This study is currently in progress~\cite{workinprogress}.

\ack
This work has been supported by Plan Nacional de Altas Energ\'{\i}as
 FPA2009-07908 and FPA2008-01430, CPAN (CSD2007-00042), Comunidad de
 Madrid HEP-HACOS S2009/ESP-1473.  The research of E.M. is supported
 by the Juan de la Cierva Program of the Spanish MICINN. L.M. has been
 supported by fellowship BES-2010-041571. F.P. has been supported by
 fellowship CPI Comunidad de Madrid.

\appendix

\section{Holographic Renormalization}
\label{sec:holo_renorm}

In order to have a deeper understanding of the holographic model, we
will compute the renormalized action within the Hamiltonian approach,
see e.g.~\cite{Martelli:2002sp,Papadimitriou:2004ap}.  Without loss of
generality we choose a gauge in which $A_r=0$ and the bulk metric
writes
\begin{equation}
 ds^2 = dr^2 + \gamma_{ij} dx^i dx^j \,.
\end{equation}
Latin letters denote four dimensional (boundary) indices. The non
vanishing Christoffel symbols are~$\Gamma^r_{ij} = -K_{ij} = -\frac 1
2 \dot{\gamma}_{ij}$ and $\Gamma^i_{jr} = K^i_j $, where dot denotes
differentiation respect $r$. Then one can compute the off shell action
in terms of transverse components of tensors. It is useful to divide
the action up in three terms: {\small\begin{eqnarray}
\label{eq:Sb1} 
S^0 &=& \frac{1}{16 \pi G} \int  d^5x\,\sqrt{-\gamma} \left[ \hat{R} + 2 \Lambda + K^2 -
K_{ij}K^{ij} - \frac 1 2 E_i E^i - \frac 1 4 \hat{F}_{ij}\hat{F}^{ij} \right] \, ,\\
\label{eq:Sb2}
S^{1}_{CS} &=& -\frac{\kappa}{12\pi G} \int d^5x\,\sqrt{-\gamma}  \epsilon^{ijkl} A_i E_j \hat{F}_{kl}  \,,\\
\label{eq:Sb3}
S^2_{CS} &=& -\frac{8 \lambda}{16\pi G} \int d^5x \sqrt{-\gamma}  \epsilon^{ijkl} \bigg[ 
A_i \hat{R}^n\,_{mkl} D_nK^m_j + E_i K_{jm}D_k K^m_l + \frac 1 2 \hat{F}_{ik} K_{jm}\dot{K}^m_l \bigg] \,. 
\end{eqnarray}}
The first one is the usual gravitational bulk and gauge terms with the
usual Gibbons-Hawking term. Of particular concern is the last term in
$S^2_{CS}$ which contains explicitly the normal derivative of the
extrinsic curvature $\dot K_{ij}$. Then the field equations will be
generically of third order in $r$-derivatives. Having applications to
holography in mind we can however impose the boundary condition that
the metric has an asymptotically AdS expansion of the form
\begin{equation}
\gamma_{ij} = \e^{2 r} \left( g^{(0)}_{ij} + e^{-2r} g^{(2)}_{ij} + e^{-4r} (g^{(4)}_{ij}
  + 2r \tilde{g}^{(4)}_{ij} )+ \cdots\right) \,, \qquad r\rightarrow\infty \,.
\end{equation}
Using this expansion one can see that the last term in the action does
not contribute in the limit $r\rightarrow\infty$. Therefore the
boundary action depends only on the boundary metric $\gamma_{ij}$ but
not on the derivative $\dot \gamma_{ij}$.

The renormalization procedure follows from an expansion of the four dimensional quantities in eigenfunctions of the dilatation operator
\begin{equation}
\delta_D = 2 \int d^4x \gamma_{ij} \frac{\delta}{\delta \gamma_{ij}} \,. 
\end{equation}
For the extrinsic curvature and gauge fields, this expansion reads respectively
\begin{eqnarray}
K^i_j &=& K_{(0)}\,^i_j + K_{(2)}\,^i_j + K_{(4)}\,^i_j + \tilde{K}_{(4)}\,^i_j \log e^{-2r} + \cdots \,, \label{eq:Kijexp} \\
A_i &=& A_{(0)}\,_i + A_{(2)}\,_i + \tilde{A}_{(2)}\,_i \log e^{-2r} + \cdots \,, \label{eq:Aiexp}
\end{eqnarray}
where the subindexes denote the corresponding eigenvalue, except for the cases~$\delta_D K_{(4)}\,^i_j = -4 K_{(4)}\,^i_j - 2 \tilde{K}_{(4)}\,^i_j $ and $\delta_D A_{(2)}\,_i  = -2 A_{(2)}\,_i - 2 \tilde{A}_{(2)}\,_i$. Given the above expansion of the fields one has to solve the equations of motion in its Codazzi form, order by order in a recursive way. The derivative on $r$ can be computed by using 
\begin{equation}
\frac{d}{dr} = \int d^4x \dot{\gamma}_{km} \frac{\delta}{\delta \gamma_{km}} = 2 \int d^4x K^l_m \gamma_{lk} \frac{\delta}{\delta \gamma_{km}}  \,. \label{eq:dr}
\end{equation}
By inserting in this equation the expansion of $K^i_j$ given by Eq.~(\ref{eq:Kijexp}), one gets $d/dr \simeq \delta_D$ at the lowest order. Taking into account this, the computation of $K_{(0)}\,^i_j$ is trivial, i.e.
\begin{equation}
K_{(0)} \,_{ij} = \frac{1}{2} \dot\gamma_{ij} \big|_{(0)} = \frac{1}{2}\delta_D \gamma_{ij} = \gamma_{ij} \,, \qquad  K_{(0)} = d \,. \label{eq:K0comp}
\end{equation}
By following the procedure and using the Codazzi form of equations of motion up to fourth order~\cite{Landsteiner:2011iq}, one gets
\begin{eqnarray}
K_{(2)} &:=& P = \frac{\hat{R}}{2(d-1)} \,, \qquad K_{(2)}\,^i_j := P^i_j = \frac{1}{(d-2)} \left[ \hat{R}^i_j - P \delta^i_j \right] \,, \label{eq:Pij} \\ 
K_{(4)} &=& \frac{1}{2(d-1)} \bigg[ P^i_j P^j_i - P^2 - \frac{1}{4} \hat{F}_{(0)}\,_{ij} \hat{F}_{(0)}\,^{ij}  \bigg] \,, \qquad \tilde{K}_{(4)} = 0 \,. \label{eq:K4tilde}
\end{eqnarray}
 
In order to compute the counterterm for the on-shell action, besides the equations of motion an additional equation is needed. Following Ref.~\cite{Papadimitriou:2004ap}, one can introduce a covariant variable $\theta$ and write the on-shell action as
\begin{equation}
S_{on-shell} = \frac{1}{8\pi G} \int_\partial d^4x \sqrt{-\gamma} (K - \theta) \,. \label{eq:Sonshelltheta}
\end{equation}
The variable $\theta$ admits also an expansion in eigenfunctions of $\delta_D$ of the form~$\theta = \theta_{(0)} + \theta_{(2)} + \theta_{(4)} + \tilde\theta_{(4)}\log e^{-2r} + \cdots $. Using the corresponding Codazzi equation for $\theta$~\cite{Landsteiner:2011iq}, and the same procedure as above,  one gets the result
\begin{eqnarray}
&&\theta_{(0)} = 1 \,, \qquad \theta_{(2)} = \frac{P}{(2-d)} \,.  \label{eq:theta02} \\
&&\tilde\theta_{(4)} = \frac{1}{4} \bigg[ P^i_j P^j_i - P^2 - \frac{1}{4}  \hat{F}_{(0)}\,_{ij} \hat{F}_{(0)}\,^{ij} + \frac{1}{3} D_i\left( D^i P - D^j P^i_j \right)  \bigg] \,. \label{eq:theta4tilde}
\end{eqnarray}
Finally the counterterm of the action can be read out from Eq.~(\ref{eq:Sonshelltheta}) by using $K$ and~$\theta$ computed up to fourth order. The result is
\begin{eqnarray}
S_{ct} &=& - \frac{(d-1)}{8\pi G} \int_\partial d^4x \sqrt{-\gamma} \bigg[
1 + \frac{1}{(d-2)}P \nonumber \\
&&\qquad\qquad- \frac{1}{4(d-1)} \left( P^i_j P^j_i - P^2 -  \frac{1}{4} \hat{F}_{(0)}\,_{ij} \hat{F}_{(0)}\,^{ij} \right)\log e^{-2r} \bigg] \,.
\end{eqnarray}
We have explicitly checked that the $\lambda$ dependence starts
contributing at sixth order. So the gauge-gravitational Chern-Simons
term does not induce new divergences, and there is no contribution in
the counterterm coming from it.

\section*{References}

\bibliography{megias}

\providecommand{\newblock}{}
\begin{thebibliography}{10}
\expandafter\ifx\csname url\endcsname\relax
  \def\url#1{{\tt #1}}\fi
\expandafter\ifx\csname urlprefix\endcsname\relax\def\urlprefix{URL }\fi
\providecommand{\eprint}[2][]{\url{#2}}

\bibitem{Adler:1969gk}
Adler S~L 1969 {\em Phys.Rev.\/} {\bf 177} 2426--2438

\bibitem{Bell:1969ts}
Bell J~S and Jackiw R 1969 {\em Nuovo Cim.\/} {\bf A60} 47--61

\bibitem{Delbourgo:1972xb}
Delbourgo R and Salam A 1972 {\em Phys.Lett.\/} {\bf B40} 381--382

\bibitem{Eguchi:1976db}
Eguchi T and Freund P~G 1976 {\em Phys.Rev.Lett.\/} {\bf 37} 1251

\bibitem{AlvarezGaume:1983ig}
Alvarez-Gaume L and Witten E 1984 {\em Nucl.Phys.\/} {\bf B234} 269

\bibitem{Fukushima:2008xe}
Fukushima K, Kharzeev D~E and Warringa H~J 2008 {\em Phys. Rev.\/} {\bf D78}
  074033 (\textit{Preprint} \eprint{0808.3382})

\bibitem{Yee:2009vw}
Yee H~U 2009 {\em JHEP\/} {\bf 11} 085 (\textit{Preprint} \eprint{0908.4189})

\bibitem{Rebhan:2009vc}
Rebhan A, Schmitt A and Stricker S~A 2010 {\em JHEP\/} {\bf 01} 026
  (\textit{Preprint} \eprint{0909.4782})

\bibitem{Erdmenger:2008rm}
Erdmenger J, Haack M, Kaminski M and Yarom A 2009 {\em JHEP\/} {\bf 01} 055
  (\textit{Preprint} \eprint{0809.2488})

\bibitem{Son:2009tf}
Son D~T and Surowka P 2009 {\em Phys. Rev. Lett.\/} {\bf 103} 191601
  (\textit{Preprint} \eprint{0906.5044})

\bibitem{Kharzeev:2009pj}
Kharzeev D~E and Warringa H~J 2009 {\em Phys. Rev.\/} {\bf D80} 034028
  (\textit{Preprint} \eprint{0907.5007})

\bibitem{Hou:2011ze}
Hou D, Liu H and Ren H~c 2011 {\em JHEP\/} {\bf 05} 046 (\textit{Preprint}
  \eprint{1103.2035})

\bibitem{Amado:2011zx}
Amado I, Landsteiner K and Pena-Benitez F 2011 {\em JHEP\/} {\bf 1105} 081
  (\textit{Preprint} \eprint{1102.4577})

\bibitem{Landsteiner:2011iq}
Landsteiner K, Megias E, Melgar L and Pena-Benitez F 2011 {\em JHEP\/} {\bf 09}
  121 (\textit{Preprint} \eprint{1107.0368})

\bibitem{Landsteiner:2011tg}
Landsteiner K, Megias E and Pena-Benitez F 2011 {\em arXiv:1110.3615
  [hep-ph]\/} (\textit{Preprint} \eprint{1110.3615})

\bibitem{Landsman:1986uw}
Landsman N and van Weert C 1987 {\em Phys.Rept.\/} {\bf 145} 141

\bibitem{Landsteiner:2011cp}
Landsteiner K, Megias E and Pena-Benitez F 2011 {\em Phys. Rev. Lett.\/} {\bf
  107} 021601 (\textit{Preprint} \eprint{1103.5006})

\bibitem{Evans:1995yz}
Evans T 1995  (\textit{Preprint} \eprint{hep-ph/9510298})

\bibitem{Moore:2010jd}
Moore G~D and Tassler M 2011 {\em JHEP\/} {\bf 1102} 105 (\textit{Preprint}
  \eprint{1011.1167})

\bibitem{Rubakov:2010qi}
Rubakov V~A 2010  (\textit{Preprint} \eprint{1005.1888})

\bibitem{Gynther:2010ed}
Gynther A, Landsteiner K, Pena-Benitez F and Rebhan A 2011 {\em JHEP\/} {\bf
  02} 110 (\textit{Preprint} \eprint{1005.2587})

\bibitem{Kimura:2011ef}
Kimura T and Nishioka T 2011  (\textit{Preprint} \eprint{1109.6331})

\bibitem{Son:2002sd}
Son D~T and Starinets A~O 2002 {\em JHEP\/} {\bf 09} 042 (\textit{Preprint}
  \eprint{hep-th/0205051})

\bibitem{Herzog:2002pc}
Herzog C~P and Son D~T 2003 {\em JHEP\/} {\bf 03} 046 (\textit{Preprint}
  \eprint{hep-th/0212072})

\bibitem{Kaminski:2009dh}
Kaminski M, Landsteiner K, Mas J, Shock J~P and Tarrio J 2010 {\em JHEP\/} {\bf
  02} 021 (\textit{Preprint} \eprint{0911.3610})

\bibitem{Amado:2009ts}
Amado I, Kaminski M and Landsteiner K 2009 {\em JHEP\/} {\bf 0905} 021
  (\textit{Preprint} \eprint{arXiv:0903.2209})

\bibitem{Vilenkin:1980ft}
Vilenkin A 1980 {\em Phys.Rev.\/} {\bf D22} 3067--3079

\bibitem{Vilenkin:1980zv}
Vilenkin A 1980 {\em Phys.Rev.\/} {\bf D21} 2260--2269

\bibitem{Neiman:2010zi}
Neiman Y and Oz Y 2011 {\em JHEP\/} {\bf 03} 023 (\textit{Preprint}
  \eprint{1011.5107})

\bibitem{Bhattacharyya:2008jc}
Bhattacharyya S, Hubeny V~E, Minwalla S and Rangamani M 2008 {\em JHEP\/} {\bf
  0802} 045 (\textit{Preprint} \eprint{0712.2456})

\bibitem{Banerjee:2008th}
Banerjee N {\em et~al.\/} 2011 {\em JHEP\/} {\bf 01} 094 (\textit{Preprint}
  \eprint{0809.2596})

\bibitem{workinprogress}
Landsteiner K, Megias E and Pena-Benitez F {\em {, work in progress}\/}

\bibitem{Martelli:2002sp}
Martelli D and Mueck W 2003 {\em Nucl.Phys.\/} {\bf B654} 248--276
  (\textit{Preprint} \eprint{hep-th/0205061})

\bibitem{Papadimitriou:2004ap}
Papadimitriou I and Skenderis K 2004   73--101 (\textit{Preprint}
  \eprint{hep-th/0404176})

\end{thebibliography}

\end{document}